\documentclass{mem}
\usepackage{natbib}\usepackage{txfonts}\usepackage{balance}
\usepackage{flushend}
\usepackage{graphicx}
\usepackage[breaklinks,pdftex]{hyperref}
\usepackage{xcolor}
\idline{75}{1}
\begin{document}
\def\teff{$T\rm_{eff }$}
\def\kms{$\mathrm {km s}^{-1}$}

\def\icarus{Icarus}
\def\jaavso{JAAVSO}
\def\pasa{PASA}
\def\na{New Astronomy}
\def\newa{New Astronomy}

\title{
Stellar Variability in Ground-Based \\
Photometric Surveys: An Overview
}

\author{
M. \,Catelan\inst{1,2}
          }

\institute{
Instituto de Astrofísica, Facultad de Física, Pontificia Universidad Católica de Chile, Av. Vicuña Mackenna 4860, 7820436 Macul, Santiago, Chile
\and 
Millennium Institute of Astrophysics, Nuncio Monse\~{n}or Sotero Sanz 100, Of. 104, Providencia, Santiago, Chile
\email{mcatelan@astro.puc.cl}\\ 
}

\authorrunning{Catelan}

\titlerunning{Ground-based photometric surveys}

\date{Received: 30 12 2022; Accepted: xx yyy 2023}

\abstract{
Time-resolved ground-based surveys in general, and photometric ones in particular, have played a crucial role in building up our knowledge of the properties, physical nature, and the very existence of the many different classes of variable stars and transient events that are currently known. Here I provide a brief overview of these developments, discussing some of the stumbling blocks that had to be overcome along the way, and others that may still hamper further progress in the area. A compilation of different types of past, present, and future surveys is also provided. 

}
\maketitle{}

\section{Prologue}\label{sec:prologue}
This paper provides a brief overview of the early developments, current state, and future of ground-based, wide-field, visual and near-infrared (IR) photometric time-series surveys. Surveys that have been carried out from space, in other wavelength regimes, and/or using spectroscopy are reviewed elsewhere in this volume \citep[see also][]{Djorgovski_2013}.

\section{The Early Years}\label{sec:early}
Since very early, humankind has demonstrated a keen interest in observing the sky~-- and particularly the {\em changing} sky. The earliest observers were awed by what was then a much darker night sky than we can find today in most places, and especially by such variable phenomena as solar and lunar eclipses and comets. Slowly, if not a rational understanding of these phenomena, variability {\em patterns} such as the phases of the Moon and the seasons started to become evident to systematic observers. The first calendars were born. Still, some sky observables, including temporary naked-eye celestial events, were commonly attributed to deities \citep[][]{Masse_1995}. Ritual sacrifices were often performed, in the hope that the gods would help ensure 
good crops, fertility, and even military success \citep[e.g.,][]{Sprajc_2017}. 

While many of these variable sky phenomena that so attracted the attention of early observers were related to Solar System objects or our own atmosphere, bona-fide extrasolar transient events, including ``guest stars'' (novae and supernovae), were also recorded early on, particularly in China \citep{Stephenson_2009}. Unfortunately, the ancient records of many of those cultures that were known to have a deep interest in Astronomy have either been lost or survived in oral tradition only \citep{Graur_2022}.

Eventually, Tycho Brahe and others demonstrated that the ``new star'' that became visible in AD\,1572, and which we now know was a type Ia supernova, did not have a significant parallax, and thus belonged in the realm of the ``fixed stars'' \citep{Stephenson_2017}. However, it was not until the late $16^{\rm th}$/early $17^{\rm th}$ century AD that David Fabricius conclusively identified the first bona-fide variable star in the night sky \citep{Zsoldos_2020}. It was listed as $o$~Ceti in Johan Bayer's {\em Uranometria} \citep{Bayer_1603}, but neither he nor Fabricius were aware that they had observed a periodic variable. In 1662, Johannes Hevelius gave it its now famous name (Mira, ``the wonderful''), and a few years later, Isma\"el Boulliau 
first measured its period. Mira would eventually turn out to be the prototype of the so-called {\em long-period variable}, or LPV, variability class \citep[e.g.,][]{Catelan_2015}.  

Between that time and until the late 1800s/early 1900s, the number of known variable stars increased slowly \citep[see Table~1.1 and Fig.~2.5 in][]{Catelan_2015}. The first known variable star catalog, put together by Edward Pigott in 1786, had a mere 12 entries, four of which were novae,\footnote{\citet{Hoffleit_1986} points out that of order 130 novae were already known at the time, most found in China, only about a dozen of which had been recorded in the West.} plus an additional 39 candidates \citep{Hoffleit_1986}. 

The discovery of Mira and other variables led to an increase in interest in stellar variability. This was also facilitated by the publication of Argelander's {\em Uranometria Nova} and {\em Bonner Durchmusterung}, as well as Hagen's {\em Atlas Stellarum Variabilium} \citep{Hoffleit_1986}. Edward C. \citeauthor{Pickering_1882}'s (\citeyear{Pickering_1882}) {\em Plan for Securing Observations of the Variable Stars} emphasized the importance of stellar variability studies, with a plea (to be repeated several times in subsequent years) for cooperation among professional and amateur astronomers alike. He urged the general public to conduct systematic observations that could be ``reduced to the same system,'' lest ``the time spent at a telescope [be] nearly wasted.''  
In his 1882 pamphlet, one will also find the following: 

\vskip 0.125cm 
\leftskip 0.25cm 
\noindent {\small {\em 
``Many ladies are interested in astronomy and own telescopes, but with two or three noteworthy exceptions their contributions to the science have been almost nothing. Many of them have the time and inclination for such work, and especially among the graduates of women's colleges are many who have had abundant training to make excellent observers.''} \citep{Pickering_1882}} 
\rightskip 0.25cm
\vskip 0.125cm 

\leftskip 0cm 
\rightskip 0cm

As director of the Harvard College Observatory (HCO), Pickering would soon start hiring the so-called
{\em Harvard Women Astronomical Computers}, to assist him in the processing of the massive amount of photographic plates that were being acquired by (mostly male) astronomers. Many of these women 
(especially W. Fleming and H. Leavitt, along with L. Ceraski in Moscow) 
would become prolific discoverers of variable stars \citep{Hoffleit_1991}.  
It was in the course of such work that Leavitt discovered the famous period-luminosity relation or {\em Leavitt Law} of classical Cepheids \citep{Leavitt_Pickering_1912}. 

It was also Pickering who started to amass the {\em Harvard College Observatory's Astronomical Photographic Glass Plate Collection}, which can arguably be considered the first systematic survey of the night sky. 
Unfortunately, due to lack of funding, Pickering could not bring his program to conclusion. In the words of Harlow \citet{Shapley_1943}, who in 1921 had succeded Pickering as HCO director, and would remain in this post through 1952: 

\vskip 0.125cm 
\leftskip 0.25cm 
\noindent {\small {\em ``Notwithstanding extensive work on thousands of faint variable stars, and on selected bright ones, no systematic survey of the light curves of all the variable stars down to some given magnitude for the whole sky~-- as contemplated forty years ago by Professor Pickering~-- was undertaken at the Harvard Observatory until 1937.''} \citep{Shapley_1943}}
\rightskip 0.25cm
\vskip 0.125cm 

\leftskip 0cm 
\rightskip 0cm

\noindent The systematic analysis of the HCO plates was eventually resumed, under the leadership of Cecilia Payne-Gaposchkin and her husband Sergei Gaposchkin, under the auspices of Harvard University's Milton Fund. Harvard's so-called {\em Milton Star Bureau} was inaugurated in 1937-1938 \citep{Shapley_1939,Shapley_1943}. The project was completed in 1953, with some delay caused by World War II \citep{Hoffleit_2000}. 

The Digital Access to a Sky Century @ Harvard (DASCH) project has been developing the tools 
to digitize and calibrate the Harvard collection of $\sim 500,\!000$ glass plates of astronomical images of the full sky,  
taken between 1885 and 1992 \citep{Grindlay_2009,Grindlay_2012,Tang_2013,Grindlay_2017}. Its latest ($6^{\rm th}$) and penultimate data release \citep{Los_2019} covers the entire northern sky in addition to Baade's Window and the Large Magellanic Cloud.
 
Similar efforts are ongoing at other institutions around the world \citep[][and references therein]{Grindlay_Griffin_2012,Griffin_2017,Sokolovsky_2018}. 
The importance of these 
efforts cannot be overemphasized: historical data can be crucial in the analysis of long-term variability phenomena, including, among many others, period changes and the secular evolution of 
different types of
pulsating stars \citep[e.g.,][and references therein]{van_Genderen_1997,Jurcsik_2012,Mukadam_2013,Rodriguez-Segovia_2022}.

\section{Technological Breakthroughs: En Route to the Modern Era}

\subsection{Hardware}\label{sec:hardware} 
Many technological breakthroughs have brought about major progress in the way ground-based photometric surveys are carried out. This includes photographic plates, photoelectric detectors, high-speed photometry, charge-coupled devices (CCDs),\footnote{
Willard S. Boyle and George E. Smith 
shared half of the 2009 Nobel Prize in Physics for their invention of the CCD. 
The other half was awarded to Charles K. Kao, 
for his work on fiber optics.} large-format CCDs, CCD arrays, near-IR detectors, large-format near-IR arrays, robotic telescopes, etc. \citep[e.g.,][]{Percy_1986,Rogalski_2012}. 

Techniques to process and analyze the acquired time-series data have also evolved enormously since the early days of visual and photographic observations. Such primitive tools as {\em flyspankers} \citep[described, for instance, in][]{Mowbray_1956}, iris photometers, and blink microscopes (also known as blink comparators) were widely used, but they implied a slow and tedious job of detecting variables and measuring how their brightness changed with time. 

\begin{figure*}[t]
\begin{center} 
\resizebox{0.85\hsize}{!}{\includegraphics[clip=true]{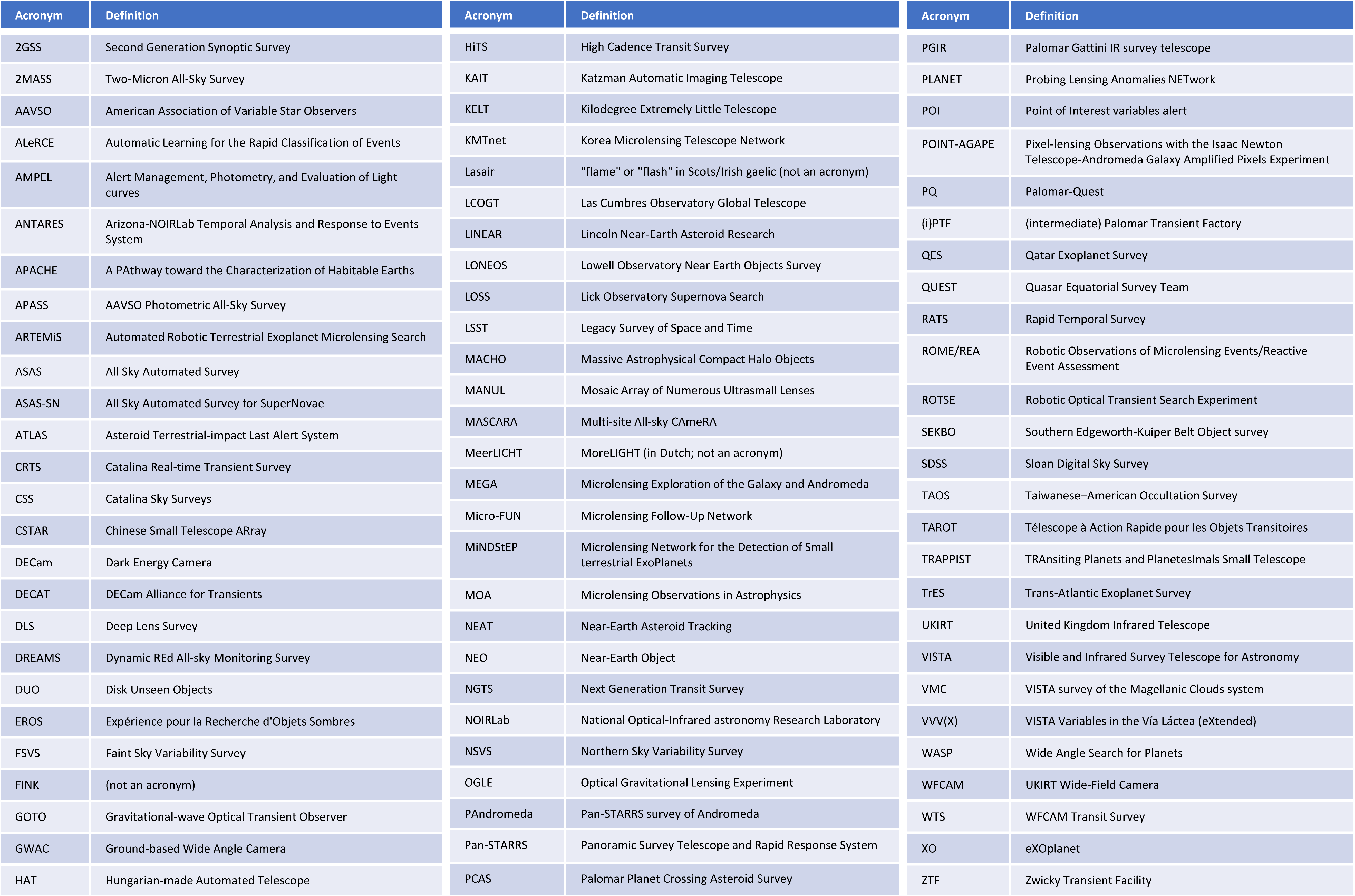}}
\caption{
\footnotesize
Non-exhaustive list of acronyms (surveys, telescopes, follow-up projects, alert brokers, etc.).}
\label{fig:acronyms}
\end{center} 
\end{figure*}

\subsection{Software}\label{sec:software} 
With the advent of electronic computers and CCDs, these tools were soon replaced by {\em software}. Semi-automated computer programs were written to perform both aperture and PSF photometry, the latter being indispensable when studying crowded fields. Packages currently in use include DAOPHOT/ALLFRAME \citep{stetson_1987,stetson_1994}, SExtractor \citep{Bertin_Arnouts_1996}, 
and DoPHOT \citep{Schechter_1993,Alonso_Garcia_2012}, 
among others
\citep[see also][for extensive references to earlier photometric packages]{Stetson_1992}. Difference imaging (or image subtraction) analysis (DIA) is another technique that gained popularity in studies of dense microlensing ($\mu$L) survey fields and the dense cores of globular clusters, where even PSF photometry may not give good results \citep[][]{Crotts_1992,Tomaney_1996}. DIA packages include ISIS \citep{Alard_1998,Alard_2000} and DanDIA \citep{Bramich_2008,Bramich_2013}, but other solutions have recently been proposed, including the GPU-accelerated PyTorchDIA \citep{Hitchcock_2021} and the Image Subtraction in Fourier Space code by \citet{Hu_2022}, both of which claim computational speeds much higher than afforded by other DIA solutions.  

Each of these techniques has its pros and cons, as far as capabilities, automation, speed, and hardware requirements. Selecting which one to use 
depends on numerous factors, such as 
field density, image size and depth, wavelength regime, speed constraints, desired photometric (and astrometric) accuracy (and precision), etc. 
\citep[see][in the case of the Vera C. Rubin Observatory's LSST]{Becker_2007}.
New packages continue to be developed, often with the specific needs of certain surveys in mind. Pipelines are in place that incorporate one or more of these techniques, and even light curve analysis tools \citep[e.g.,][and references therein]{Schlafly_2018,Sokolovsky_2018,Brennan_2022}. 

High-precision astrometry, in combination with sufficient time coverage, allows the precise determination of proper motions, as done in the case of the VISTA surveys of the Magellanic Clouds \citep[VMC;][]{Cioni_2014} and Galactic bulge/inner disk \citep[VVV;][]{Contreras_2017,Smith_2018}.
This can be extremely useful in studying the structure, dynamical evolution of and interaction between nearby galaxies and the Milky Way, as well as their respective stellar populations \citep[e.g.,][]{Braga_2018,Schmidt_2022}.

\section{The Era of Wide-Field Surveys}\label{sec:widefield}

\subsection{The Search for Microlensing ($\mu$L)}\label{sec:muL} 
The modern era of wide-field surveys started with a theoretical paper by Bohdan \citet{Paczynski_1986}, who proposed monitoring campaigns of the Magellanic Clouds to detect gravitational $\mu$L events caused by dark matter (DM) ``clumps,'' 
or 
massive astrophysical compact halo objects \citep[MACHOs;][]{Griest_1991}.  \citet{Paczynski_1991} and \citet{Griest_etal_1991} next pointed out that $\mu$L searches of the Galactic bulge could provide additional constraints on the DM content of the disk and the bottom end of the initial mass function. It was also noted that the technique could be used to detect extrasolar planets as well \citep{Mao_1991,Gould_1992}, a technique that has continuously been used since \citep[for reviews, see][]{Gaudi_2012,Tsapras_2018,Zhu_2021}.
Before long, several groups 
put together $\mu$L search campaigns, including  
MACHO \citep{Alcock_1992}, 
OGLE \citep{Udalski_1992}, 
EROS \citep{Aubourg_1993a}, 
DUO \citep{Alard_1997}, 
and MOA \citep{Abe_1997}.\footnote{Figure~\ref{fig:acronyms} provides a list of acronyms.} 
Other groups have chosen M31 as target, including MEGA \citep{Crotts_2001}, PAndromeda \citep{Lee_2012}, POINT-AGAPE \citep{Paulin_2003}, and others \citep[for a review, see][]{Lee_2016}. 

The first $\mu$L events were soon reported 
\citep{Alcock_1993,Udalski_1993,Aubourg_1993b}, and many more were to come~-- but the DM content of the Milky Way has remained elusive. On the other hand, 
it did not take long for the first {\em variable star studies and catalogs} based on these data to appear in the literature \citep[e.g.,][]{Udalski_1994,Beaulieu_1995,An_2004}.

We have gone a long way since these early experiments. Some still relied, at least in part, on photographic plates. Others were equipped with small CCDs. Some were relatively short-lived,  
while others have thrived. 
OGLE, in particular, has continued to monitor the Magellanic Clouds and Galactic bulge since its early years of operation, with each successive phase  \citep[currently the fouth, thus OGLE-IV;][]{Soszynski_2012} boasting increased sky coverage and improved hardware. OGLE has proved to be a phenomenal resource for variable star studies. Its data have led to the discovery of new types of variables, including the so-called {\em blue large-amplitude pulsators} \citep[BLAPs;][]{Pietrukowicz_2017} and {\em binary evolution pulsators}  \citep{Pietrzynski_2012,Soszynski_2016}, as well as type I \citep{Soszynski_2008} and II \citep{Soszynski_2010} Cepheids and $\delta$~Scutis \citep{Soszynski_2021} in eclipsing binary systems. OGLE discovered the first known multi-mode anomalous Cepheid \citep{Soszynski_2020b}, and witnessed a double-mode RR Lyrae star (RRL) turn into a fundamental-mode one 
in the course of 
a few years \citep{Soszynski_2014}. It has also shed light on the distribution of dual- and multi-mode classical pulsators in the Petersen diagram \citep[e.g.,][]{Soszynski_2020a,Smolec_2018,Smolec_2022}. In addition, OGLE data 
have proved instrumental in discovering some of the first strong candidate binary RRL 
\citep{Hajdu_2015,Prudil_2019}, and 
statistical studies of their mass distribution are now starting to become a reality \citep{Hajdu_2021}. 

\subsection{Going All-Sky: Transits, Transients, NEOs, etc.}\label{sec:allsky}
$\mu$L searches have focused mainly on regions of high stellar (background) density, and so covered relatively limited patches of the sky. \citet{Paczynski_1997,Paczynski_2000} advocated 
for {\em all-sky} surveys, whereas  \citet{Nemiroff_1999,Nemiroff_2003} pointed out the benefits of obtaining a {\em continuous} record of the sky {\em ``and plac[ing] it on the World Wide Web for anyone in the world.''} 
Both noted that many different science cases, in addition to $\mu$L, could be explored with such data. 
\citet{Paczynski_2000,Paczynski_2006} estimated that of order $10^6$ 
variable stars could be discovered even with small-aperture telescopes, ``and many more with larger telescopes.'' 

Since then,\footnote{An earlier effort was the PCAS photographic NEO survey \citep{Helin_1979}, conducted between 1973 and 1978, which covered 80,570~deg$^2$
with the Palomar 46\,cm Schmidt camera.}   all-sky or synoptic surveys, many of which use 
small telescopes (located at a specific site or distributed around the world), have become more the norm than the exception. Each 
has focused on one or more of the 
science goals laid out by the pioneers, such as:

\begin{itemize}
 \item[$\bullet$] {\em\textbf{Planetary transits}}:\\ 
    HAT \citep{Bakos_2002}, 
    KELT \citep{Pepper_2007}, 
    KMTNet \citep{Kim_2016}, 
    MASCARA \citep{Snellen_2012}, 
    NGTS \citep{Wheatley_2018}, 
    QES \citep{Alsubai_2013}, 
    Solaris \citep{Kozlowski_2017}, 
    TRAPPIST \citep{Gillon_2011}, 
    TrES \citep{Alonso_2004}, 
    WASP and SuperWASP \citep{Pollacco_2006}, 
    XO \citep{McCullough_2005}, 
    etc.;  

\vskip0.125cm
 \item[$\bullet$] {\em\textbf{Transient events}}: \\
    ASAS-SN \citep{Kochanek_2017}, 
    CRTS \citep{Drake_2009}, 
    HiTS \citep{Forster_2016}, 
    LOSS \citep{Filippenko_2001}, 
    Palomar-Quest \citep{Djorgovski_2008}, 
    PTF \citep{Law_2009} and iPTF \citep{Kulkarni_2013}, 
    ROTSE \citep{Akerlof_2000}, 
    SuperMACHO \citep{Becker_2005}, 
    ZTF \citep{Bellm_2019}, 
    etc.; 
 
\vskip0.125cm
 \item[$\bullet$] {\em\textbf{Near-Earth/Solar System objects, ``killer'' or ``doomsday'' asteroids}}:\\
    ATLAS \citep{Tonry_2018},
    CSS \citep{Larson_1998,Seaman_2022}, 
    ESA's Flyeye \citep{Perozzi_2021}, 
    LINEAR \citep{Stokes_2000}, 
    LONEOS \citep{Wagner_1998}, 
    NEAT \citep{Helin_1997}, 
    Pan-STARRS \citep[][]{Kaiser_2002},\footnote{Pan-STARRS is actually a survey {\em telescope} that has been used to carry out different types of surveys \citep[see][]{Chambers_2016}.}
    etc. 
\end{itemize}

In addition to these, \textbf{\em multi-purpose wide-field surveys} have also been carried out 
or will soon see first light. 
Many of these incorporate specific classes of variable stars
among their primary goals. Examples include 
AAVSOnet \citep{Simonsen_2012}, 
APASS and 2GSS \citep{Henden_2017}, 
ASAS \citep{Pojmanski_1997}, 
CSTAR \citep{Yuan_2008}, 
DECam DDF programs run under the DECAT alliance \citep{Graham_2023}, 
Evryscope \citep{Law_2015}, 
Hungarian Fly's Eye \citep{Pal_2013,Meszaros_2019}, 
FSVS \citep{Groot_2003}, 
MeerLICHT and BlackGEM \citep{Bloemen_2016}, 
NSVS \citep{Wozniak_2004}, and, of course, Rubin/LSST (\citeauthor{Ivezic_2019} \citeyear{Ivezic_2019}; see also \citeauthor{Hambleton_2022} \citeyear{Hambleton_2022}, specifically in the context of transients and variable stars). 
Some surveys have also targeted \textbf{\em specific classes or subclasses of variable stars}, such as the SEKBO \citep{Keller_2008} and QUEST \citep{Vivas_2004} RRL surveys, as well as the RATS \citep{Ramsay_2005} and OmegaWhite \citep{Macfarlane_2015} surveys of short-period variables. Countless time-series ``mini surveys'' covering smaller fields have been carried out as well. 

There are also some \textbf{\em projects whose main science goals may not include the time domain, but that may have secured (sometimes extensive) time-series data}, with the purpose of calibration and standardization and/or to obtain deeper co-adds. 
Examples include 2MASS \citep[][]{Skrutskie_2006}, UKIRT/WFCAM \citep[][and references therein]{Hodgkin_2009,Leggett_2020}, SkyMapper's Southern Survey \citep{Keller_2007,Onken_2019}, and SDSS's Stripe 82 \citep[][]{Ivezic_2007}. Naturally, such data can also be valuable for stellar variability studies \citep[e.g.,][]{Sesar_2007,Cross_2009,Quillen_2014,Ferreira_Lopes_2015};  
in fact, the discovery of variability in both H-deficient and ZZ~Ceti (DAV) stars was an offshoot of work on standard stars \citet{Landolt_2007}.

As in the case of OGLE, WASP, and PTF, different phases of some of these surveys have been 
carried out, are being implemented, and/or future upgrades have been proposed. Those are not listed here for lack of space.

Lastly, we note that there are many complementary and collaborative \textbf{\em follow-up networks} in place, such as PLANET \citep{Albrow_1998}, ARTEMiS \citep{Dominik_2008}, and RoboNet-II \citep{Tsapras_2009}, among others \citep[see][for a review]{Perryman_2018}. They have made extensive use of small- and medium-size telescopes around the world, such as those belonging to the LCOGT network \citep{Shporer_2011}, and new actors are entering the field \citep[][and references therein]{Han_2021,Hoffmann_2022}. ROME/REA \citep{Tsapras_2019} is an example of a $\mu$L exoplanet search-plus-follow-up project that was specifically {\em designed} having such facilities in mind. These networks are becoming increasingly important with the rise of {\em multi-messenger astronomy} \citep[][]{Branchesi_2016,Dyer_2022}.

\subsection{The Near-IR Domain}\label{sec:nir} 
Time-resolved near-IR surveys of the inner Milky Way, such as VVV(X) \citep{DMinniti_2010,DMinniti_2018} and the IRSF/SIRIUS survey by \citet{Matsunaga_2009}, have played instrumental roles in probing its structure and evolution, by piercing through large columns of foreground extinction that hamper analysis in the visual. VVV(X) in particular has covered a wide sky area around the Galactic bulge and inner disk, and led to the discovery of millions of variable star candidates \citep{Ferreira_Lopes_2020,Molnar_2022}. VMC \citep{Cioni_2011} has used the same VISTA 4\,m telescope \citep{Sutherland_2015} as VVV(X), equipped with the same near-IR camera \citep[VIRCAM;][]{Dalton_2006}. VIRCAM is comprised of an array of 16 $2048 \times 2048$ IR detectors, covering the spectral range $0.9\! - \! 2.5\,\mu{\rm m}$. At VISTA's $f/3.25$ Cassegrain focus, which affords a $1.65^{\circ}$ diameter field of view (FOV), this gives a pixel scale of $0.34''$/pixel. These specifications are unprecedented for a near-IR instrument. 

The main focus of both VVV(X) and VMC has been on variable stars. With different goals in mind, other time-series near-IR surveys covering relatively large fields on large telescopes have also been carried out. An example is WTS \citep{Kovacs_2013}, which used the WFCAM camera at the UKIRT telescope \citep{Casali_2007} to detect exoplanetary transits around M stars. Four fields were extensively observed in $J$, producing hundreds of images covering a timespan of several years that have also proved useful for stellar variability studies \citep{Birkby_2012,Nefs_2012}. Also worthy of notice are the UKIRT microlensing surveys \citep{Shvartzvald_2017}, which acquired hundreds of $H$ and $K$ images of the inner Galactic bulge between 2015 and 2019, but remain a virtually untapped resource, as far as stellar variability studies go.

\begin{figure*}[h]
\begin{center} 
\resizebox{0.8\hsize}{!}{\includegraphics[clip=true]{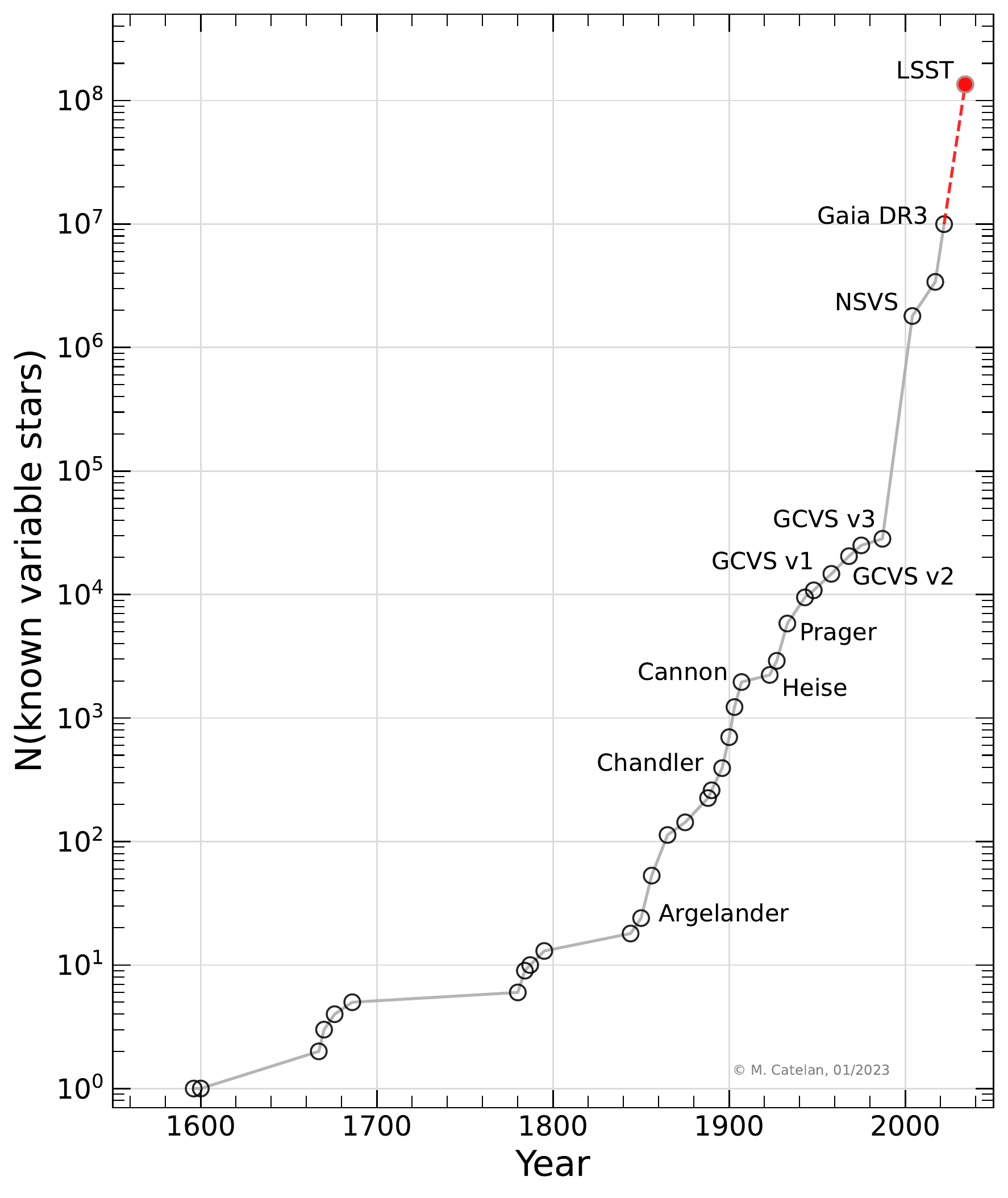}}
\caption{
\footnotesize
Number of known variable stars as a function of time. The last datapoint (in red) corresponds to the estimated number by the end of the main Rubin/LSST survey \citep{LSST_2009}.}
\label{fig:Nvar}
\end{center} 
\end{figure*}

\section{Challenges}\label{sec:challenges} 

\subsection{Near-IR Surveys}\label{sec:nirchallenges}
Unfortunately, VIRCAM will soon be retired  \citep{Bellido_2022},\footnote{4MOST, VIRCAM's replacement at the VISTA telescope, is also a unique instrument, but its goal is to perform wide-field, multi-object spectroscopy.}
leaving the community without comparable 
facilities to conduct wide-field near-IR photometric surveys of the southern sky from the ground. 
WFCAM@UKIRT \citep{Casali_2007} is not optimally placed to probe the southern sky; compared with VIRCAM (\S\ref{sec:nir}), it also has fewer detectors and covers a smaller area of the sky at a slightly poorer pixel scale. On the other hand, small-diameter robotic telescopes are now starting to be used to produce truly wide-field imaging in the near-IR. PGIR, a pathfinder instrument for near-IR time-domain astronomy, uses a 30\,cm robotic telescope with a FOV of 25~square degrees \citep{De_2020}. Other efforts in this direction include WINTER \citep{Lourie_2020}, which is located (like PGIR) at Palomar Observatory but uses a 1\,m telescope, and DREAMS \citep{Soon_2018}, at Siding Spring Observatory.

\subsection{Curating the New Variable Stars}\label{sec:curation}

There is a natural synergy between the 
many surveys mentioned in the previous section and variable star research. This is reviewed in some detail by \citet[][]{Soszynski_2017}, in the case of $\mu$L, and  \citet[][]{Kovacs_2017}, in the case of transiting exoplanets. Indeed, our current knowledge of stellar variability across the Milky Way and beyond has been built in part by some of these projects. Many of them, as we have seen, were designed with different (main) scientific goals. Properly curating the ever-increasing number of variable stars is a growing challenge. 

Since after World War II,
official IAU IDs of newly discovered variables have been assigned and listed in the General Catalog of Variable Stars (GCVS), currently in version 5.1 \citep{Samus_2017}. 
New GCVS name lists are periodically published, the latest one being the $84^{\rm th}$ \citep{Samus_2021a}. As of this writing (Dec. 2022), GCVS includes a total of 880,606 variables. Variable stars in globular clusters are now included as well \citep{Samus_2009,Samus_2022}. The AAVSO Variable Star Index (VSX; \citeauthor{Watson_2006} \citeyear{Watson_2006}; \citeauthor{Williams_2011} \citeyear{Williams_2011}), in addition to the named GCVS variables, incorporates variable star catalogs that have been published by different teams. The differences in philosophy between GCVS and VSX are laid out and discussed in \citet{Samus_2017}. 
As of this writing, VSX includes 2,206,649 variables, plus $\sim\!\!30,\!000$ candidates. 
As shown in Figure~\ref{fig:Nvar}, the number of known variables is currently $\sim\!\!10^7$, and it will likely reach $\sim\!\!10^8$ by the mid-2030s \citep[][]{Ivezic_2007}. This places an enormous challenge to careful curation efforts as done by the GCVS team. Discussing the future of variable star catalogs in the era of wide-field/high-precision ground- and space-based 
surveys, \citet{Samus_2021b} offer the following insight: 

\vskip 0.125cm 
\leftskip 0.25cm 
\noindent {\small {\em ``There is no doubt that the history of traditional variable star catalogues is approaching its end. However, it is of 
[the utmost]
importance that future astronomers do not begin to discover the same variable star again and again. A possible way out of this problem is to include variable-star information in major star catalogues of general purpose, like it has already been done in the Gaia catalogue.''} \citep{Samus_2021b}}
\rightskip 0.25cm
\vskip 0.125cm 

\leftskip 0cm 
\rightskip 0cm

\subsection{Calibration and Standardization}\label{sec:calibration} 
The study of variable stars and transients can greatly benefit from both new and historical photometric data (\S\ref{sec:early}). In the visual, newer data are collected increasingly more often in the SDSS or similar systems, whereas the 
Johnson-Kron-Cousins system prevailed until recently. Even older data will be in some visual or photographic system (if any). Putting together data from such heterogeneous sources can constitute a very serious challenge. 

\citet{Johnson_1952} pointed out early on that {\em ``transformations between different color or magnitude systems are, in general, nonlinear.''}
\citet{Johnson_1963} also noted that the widely used 
blue/yellow 
systems \citep[][]{Landolt_2016} could not be reliably defined {\em ``by visual and photographic means... because of difficulties with the calibration and standardization of visual and photographic receivers.''} He advocated for the development of an accurate, standard photometric system, pointing out that the {\em ``development and mass production of good photomultipliers''} had made this feasible. A few years earlier, these ideas had led to the definition of the first modern photometric systems, including, of course, the $UBV$ system of \citet{Johnson_1953}, among many others. 

Great effort has been devoted to the crucial task of accurately calibrating and establishing all-sky standards for these systems \citep{Bessell_2005,Landolt_2016}. Dealing with the $U$ band has proven especially problematic \citep[][]{Stetson_2000,Bessell_2005,Stetson_2019}. 
Apart from this, the same names are sometimes assigned to bandpasses in different systems, as happens, for instance, with the \citet{Cousins_1976} $VRI$ and \citet{Kron_1953} $RI$ systems. This can give rise to confusion, as they have different response curves and transformations between them can even be nonlinear \citep[][and references therein]{Bessell_1987}.\footnote{In the case of single-band surveys, it is not possible to reliably transform the data to other systems, as no color terms can be properly computed.} 
A given system can itself evolve over time \citep{Johnson_1963}. 
To complicate matters, foreground extinction will affect filters with even slightly different response curves in different ways. 

Photometric systems abound \citep[e.g.,][]{Johnson_1963,Fiorucci_2003,Bessell_2005}, and new ones continue to appear in the literature. 
When bands with similar names exist, a subscript can be used to distinguish them: examples include $I_C$ \citep{Cousins_1976} and $K_s$ \citep{Skrutskie_2006}. Primes are also used, as in the case of SDSS $u'g'r'i'z'$ \citep{Fukugita_1996}, to be compared with, say, SkyMapper's $uvgriz$ \citep{Bessell_2011}. However, these markers are not universally adopted. 
When using historical data and/or data from heterogeneous sources, it is thus important to ascertain what system or systems may have been used by different authors, and whether the data can be reliably brought to a common system.\footnote{This should also be kept in mind when comparing the data with theoretical models, as the latter's output may be based on transmission curves that do not match those used to acquire the empirical data.} 
While many recipes exist for transforming data from one system to another \citep[e.g.,][]{Jordi_2006,Pancino_2022}, 
it should be kept in mind that this too is a complex task involving multiple steps that is subject to potential pitfalls \citep[e.g.,][]{Roberts_1995,Landolt_2007,Hajdu_2020,Pancino_2022}. 

Last but not least, accurately recording the {\em times} when observations are taken is a crucial task. When combining data from heterogeneous sources, the end user should always check what {\em kind} of time may have been recorded in each case \citep[see, for instance,][for a useful summary]{McCarthy_2011}, and, if necessary, carefully bring those into a consistent time reference frame. The importance of doing this, especially in the case of long-term variability studies, cannot be overemphasized \citep{Stephenson_2005,Sterken_2005}.

\subsection{Data Tsunami}\label{sec:tsunami} 

The data volume generated by surveys has been increasing exponentially over time \citep{Djorgovski_2013,Djorgovski_2022}. The ``data explosion'' of the 1990s, when datasets were measured in GB and TB, is giving way to a veritable ``data tsunami,'' with data volumes at the several TB {\em per night} level, and full survey data comprising dozens of PB \citep[Rubin/LSST;][]{Graham_2019,Ivezic_2019}. 
Traditional methods of data management, processing, analysis, and archiving are quickly becoming infeasible. For obvious reasons, the problem is especially severe in the case of time-series studies. New techniques are emerging, with astroinformatics and artificial intelligence playing prominent roles in this new ``big data'' era \citep{Djorgovski_2022}.  

Classification of transients, variable stars, and other types of variable sources is increasingly being performed using machine learning (ML) and/or deep learning (DL) tools. This often needs to be done in real time, as in the case of the so-called {\em alert brokers} \citep[see][for a list]{Hambleton_2022}, so that time-critical events of special astrophysical interest can be quickly told apart from bogus ones, classified, and followed up. 
Images \citep[e.g.,][]{Carrasco-Davis_2019,Gomez_2020}, photometry \citep[e.g.,][]{Sanchez_2021,Narayan_2018}, or both \citep[e.g.,][]{Forster_2021} have been used, as have light curves transformed {\em into} images \citep{Szklenar_2022} and information from miscellaneous catalogs from the literature.

Different approaches to the problem have been explored in the literature, including supervised, semi-supervised, and unsupervised techniques, depending on whether a training set exists, how much data it may contain, and how representative it may be. 
Whatever the technique of choice, one must be aware of the different sources of problems that may adversely affect the results. Computational speed must also be carefully considered; for instance, depending on the technique, periods can be very slow to compute. Feature selection and design, scaling, treatment of errors in the data, data heterogeneity, different types of biases (affecting, e.g., the training sets, and even the experts themselves), and others \citep[e.g.,][and references therein]{Pantoja_2022} are all aspects that must be considered when implementing a classifier {\em and}  interpreting its results. 

Whatever the approach, the community should intensify its efforts to develop {\em multi-band time-series analysis tools and classifiers}, so that the maximum amount of information can be extracted from multi-band, time-resolved data, when available~-- as will be the case, in particular, with Rubin/LSST.

\subsection{Data Archiving and the Virtual 
Observatory}\label{sec:archiving} 

The surveys described in the previous section (and many others not listed) provide a wealth of time-series data that could be explored in studies of stellar variability and transients. It would be extremely helpful if the time-domain astronomer could quickly query all of these surveys for data that may be available for a certain star, group of stars, area of the sky, etc. Yet, combining data from even a small number of surveys is often a difficult task. 

There are many reasons for this. Cross-matching data from different catalogs with different astrometric accuracies can easily lead to mismatches, particularly in dense fields. Some groups have not made their data accessible through permanent, public links, nor followed the ``findable, accessible, interoperable, and reusable'' (FAIR) principles for data management and stewardship \citep{Wilkinson_2016}. Adopting the FAIR approach can make one's data much easier to remotely access and analyze using Virtual Observatory-enabled tools as Aladin \citep{Boch_2011}, TOPCAT \citep{Taylor_2005}, and others \citep[e.g.,][]{Araya_2015}, thus fostering scientific progress while at the same time increasing the visibility of one's survey.

\subsection{Light Pollution: Ground and Space}\label{sec:pollution}

Light pollution from ground and space sources were recently reviewed by \citet{Green_2022}. Here we focus on recent developments involving the so-called {\em satellite constellations}. 

Bright satellite trails caused by the latter are now well documented \citep[e.g.,][]{Mroz_2022}. They become more damaging in the case of high-altitude satellites, long exposures, and/or during and just after/before twilight, and are especially detrimental to wide-field images  \citep{Hainaut_2020,Tyson_2020, Bassa_2022}. While twilight images are most affected, \citet{Bassa_2022} have shown that, at Cerro Paranal, hundreds of satellites may remain visible during dark time, including dozens above $30^{\circ}$ elevation. False transient alerts may be triggered by glints produced by satellites and other space junk in Earth's orbit \citep{Karpov_2022}. \citet{Groot_2022} has pointed out that even the short-duration occultations that are caused by these objects as they cross the optical path towards distant astronomical sources may become relevant, particularly in the case of time-series data obtained with the next generation of high-speed sCMOS detectors \citep[see, e.g.,][]{Karpov_2019},

Mitigation strategies have been devised and implemented by some companies, including coating the satellites and using sun-blocking shades, while others are in discussions with the IAU and other astronomical organizations to find ways to reduce their negative impact \citep{Witze_2022}. There is, however, no legal mechanism in place 
to {\em enforce} any of this. In the words of \citet{Green_2022}, 

\vskip 0.125cm 
\leftskip 0.25cm 
\noindent {\small {\em ``unlike sources of ground-based light pollution requiring cooperation and control by local regulation, limiting the strongly negative impact of satellite constellations requires cooperation and possible regulation at national and
international level.''} \citet{Green_2022}}
\rightskip 0.25cm
\vskip 0.125cm 

\leftskip 0cm 
\rightskip 0cm

\noindent Implementing active avoidance when scheduling observations may be helpful \citep{Hu_2022b}, but this may be hampered by the lack of (or insufficiently accurate) published ephemerides \citep{Cui_2022}.

In the meantime, larger and brighter satellite constellations are coming to life. BlueWalker~3 prototype's antenna is ``the size of a squash court'' \citep{OCallaghan_2022}. Over a hundred of these are expected to be launched by 2024, many of which will be even bigger, potentially outshining ``everything in the night sky except for the Moon'' \citep{OCallaghan_2022}.
This is unfortunately supported by recent reports that the prototype (launched on Sept. 10, 2022) became ``brighter than 99.8\% of all visible stars'' \citep{Wilkins_2022}, and measurements indicating that it reached between +0.0 to +1.5~mag, once its giant flat-panel antenna array was unfolded \citep{Mallama_2022,Mallama_2023}. 
This should be compared with values between +4.0 and +7.6 for Starlink and OneWeb satellites \citep[][]{Bassa_2022}. 

The future of ground-based wide-field photometric surveys, and indeed of ground-based astronomy as a whole, is clearly in great peril. An entire discipline faces an existential threat. Calls for our community to support the {\em space environmentalist} movement \citep{Lawrence_2022} should be heeded, before it is too late.

\begin{acknowledgements}
I warmly thank Horace A. Smith and Igor Sozy\'{n}sky for useful discussions and information, and the referee for some helpful suggestions. I also gratefully acknowledge the support provided by ANID's Millennium Science Initiative through grant ICN12\textunderscore 12009, awarded to the Millennium Institute of Astrophysics (MAS), and by ANID's Basal project FB210003. 
\end{acknowledgements}

\bibliography{Catelan_bibliography_2022.bib}
\bibliographystyle{aa}

\end{document}